\documentclass[3p,times,twocolumn]{elsarticle}

\usepackage{ecrc}

\volume{00}

\firstpage{1}

\journalname{Nuclear Physics B Proceedings Supplement}

\runauth{}

\jid{nuphbp}

\jnltitlelogo{Nuclear Physics B Proceedings Supplement}

\usepackage{amssymb}

\usepackage{lineno}

\usepackage[figuresright]{rotating}

\usepackage{graphicx}
\usepackage{dcolumn}
\usepackage{bm}
\usepackage{amssymb}
\usepackage{xspace}

\usepackage{hyperref}
\usepackage{amsmath}

\newcommand{\pbpb}{\ensuremath{\rm Pb\!-\!Pb}\xspace}
\newcommand{\pp}{pp\xspace}
\newcommand{\ppb}{p-Pb\xspace}
\newcommand{\pt}{\ensuremath{p_{\text{T}}}\xspace}

\usepackage{color}

\begin{document}

\begin{frontmatter}



\dochead{}

\title{Overview of ALICE results}


\author{Antonio Ortiz Velasquez \\ on behalf of the ALICE Collaboration}

\address{Instituto de Ciencias Nucleares, Universidad Nacional Aut\'onoma de M\'exico. \\ Circuito exterior s/n, Ciudad Universitaria, Del. Coyoac\'an, C.P. 04510, M\'exico DF.}

\begin{abstract}
The ALICE detector was designed to study the physics of matter under extreme conditions of high energy density. Different results were reported by the experiment using data from the successful run I of the LHC. The goal of the present work is to present an overview of recent ALICE results. This comprises selected results from several analyses of \pp, \ppb and \pbpb data at the LHC energies. 
\end{abstract}


\end{frontmatter}


\section{Introduction}
\label{intro}

Matter which surrounds us can be found in a variety of phases, the changes on its external conditions allows to go from one to another phase. From lattice QCD, it is predicted that hadronic matter under extreme conditions of high temperature and density changes its properties, in this state the fundamental degrees of freedom are given by quarks and gluons~\cite{BraunMunzinger:2008tz}. The phase transition occurs at a critical (crossover) temperature between 143-171 MeV~\cite{Borsanyi:2010bp,PhysRevD.71.034504,PhysRevD.74.054507,PhysRevLett.113.082001,PhysRevD.85.054503,Ayala:2014jla}.  The experimental access to explore and test the QCD phase diagram and to address the fundamental question of hadron confinement and chiral
symmetry breaking is via ultra-relativistic heavy ion collisions. 

The hot and dense matter has been studied in the Relativistic Heavy Ion Collider (RHIC) at Brookhaven National Lab, the collisions of Au-Au at $\sqrt{s_{\rm NN}}=$ 0.2 TeV  produced  a perfect fluid that is initially closer to the ideal hydrodynamic limit, a theoretical scenario in which the viscosity is zero. This result has been confirmed by experiments at the Large Hadron Collider, where \pbpb collisions at $\sqrt{s_{\rm NN}}=$ 2.76 TeV have been achieved~\cite{Heinz:2011kt}. ALICE (A Large Ion Collider Experiment) is the only experiment at the LHC designed for this purpose\cite{0954-3899-32-10-001}. It has collected data which correspond to an integrated luminosity of $\approx$10$\mu$b$^{-1}$ and $\approx$0.1nb$^{-1}$ during the successful runs of 2010 and 2011, respectively. For the hot nuclear matter effects to be studied, analyses of \pp and \ppb data were also performed. Unexpectedly, these data revealed intriguing effects which are not well understood. In this paper, a review of selected results of ALICE will be presented, this comprises several measurements in different colliding systems, \pp, \ppb and \pbpb.

\begin{figure*}[htbp]
\begin{center}
\includegraphics[keepaspectratio, width=1.0\columnwidth]{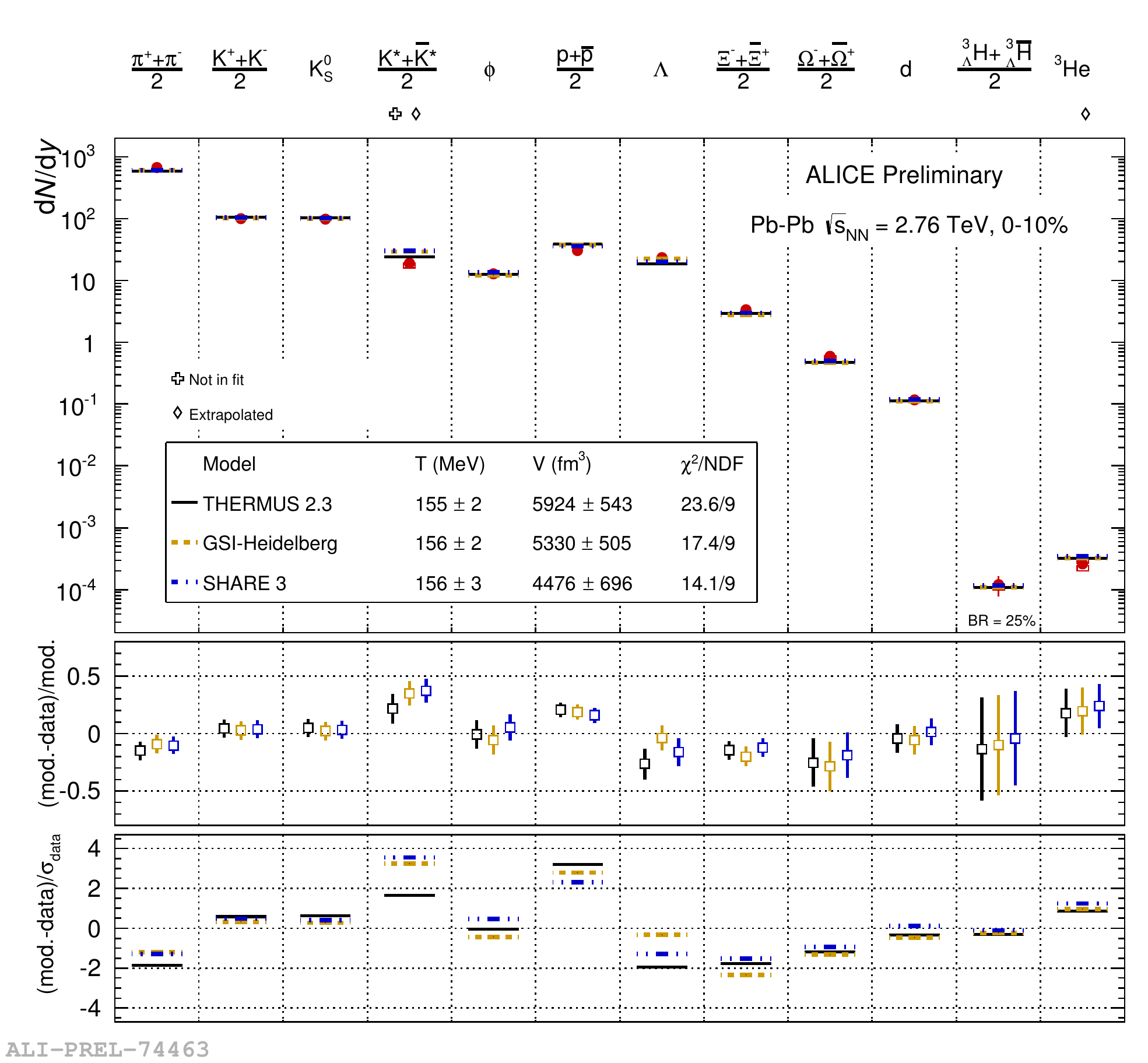}
\caption{\label{fig:pbpb:0} Thermal fits of particle yields in 0-10\% \pbpb collisions.}
\end{center}
\end{figure*}

\section{The ALICE apparatus}
\label{alice}

Particle identification (PID) is an important tool to study the hot and dense matter created in
relativistic heavy ion collisions. That is why ALICE is an experiment specialized in PID from hundreds of MeV/$c$ up to tens of GeV/$c$. The central barrel of ALICE is placed inside a large solenoidal magnet which provides a magnetic field of 0.5 T. It is dedicated to detect hadrons, electrons, and photons produced at mid-pseudorapidity, $|\eta|<$0.8. It comprises an Inner Tracking System (ITS) of high-resolution silicon detectors, a cylindrical Time-Projection Chamber (TPC), and particle identification arrays of Transition-Radiation Detectors (TRD) and of Time-Of-Flight (TOF) counters. Additional central subsystems, not-covering full azimuth, are a ring-imaging Cherenkov detector for High-Momentum Particle IDentification (HMPID), and two electromagnetic calorimeters: a high-resolution PHOton Spectrometer (PHOS) and a larger-acceptance ElectroMagnetic Calorimeter (EMCal). The muon arm detects muons emitted within  $2.5< \eta <4$ and consists of a complex arrangement of absorbers, a dipole magnet, five pairs of tracking chambers, and two trigger stations. Several smaller detectors (VZERO, TZERO, FMD, ZDC, and PMD) for triggering, multiplicity measurements and centrality determination are installed in the forward region.

\section{Latest progress on the study of the hot and dense matter}
\label{pbpb}

In central heavy ion collisions at ultra relativistic energies it is well established that a strongly interacting medium of quarks and gluons is created. To learn about the early state of the system, low \pt ($<2.2$ GeV/$c$) direct photons are used. A temperature of 304 MeV has been measured for the 0-40\% \pbpb collisions~\cite{Wilde:2012wc}. Hence, the system at the LHC is hotter than that produced at RHIC, where an early temperature of 221 MeV was measured  for the 0-20\% Au-Au collisions. The system created at the LHC is also denser, the average multiplicity per number of participant is two times that measured at RHIC~\cite{Aamodt:2010cz}. Also interesting is the fact  that the same centrality dependence is found at RHIC and LHC.

\begin{figure*}[htbp]
\begin{center}
\includegraphics[keepaspectratio, width=1.9\columnwidth]{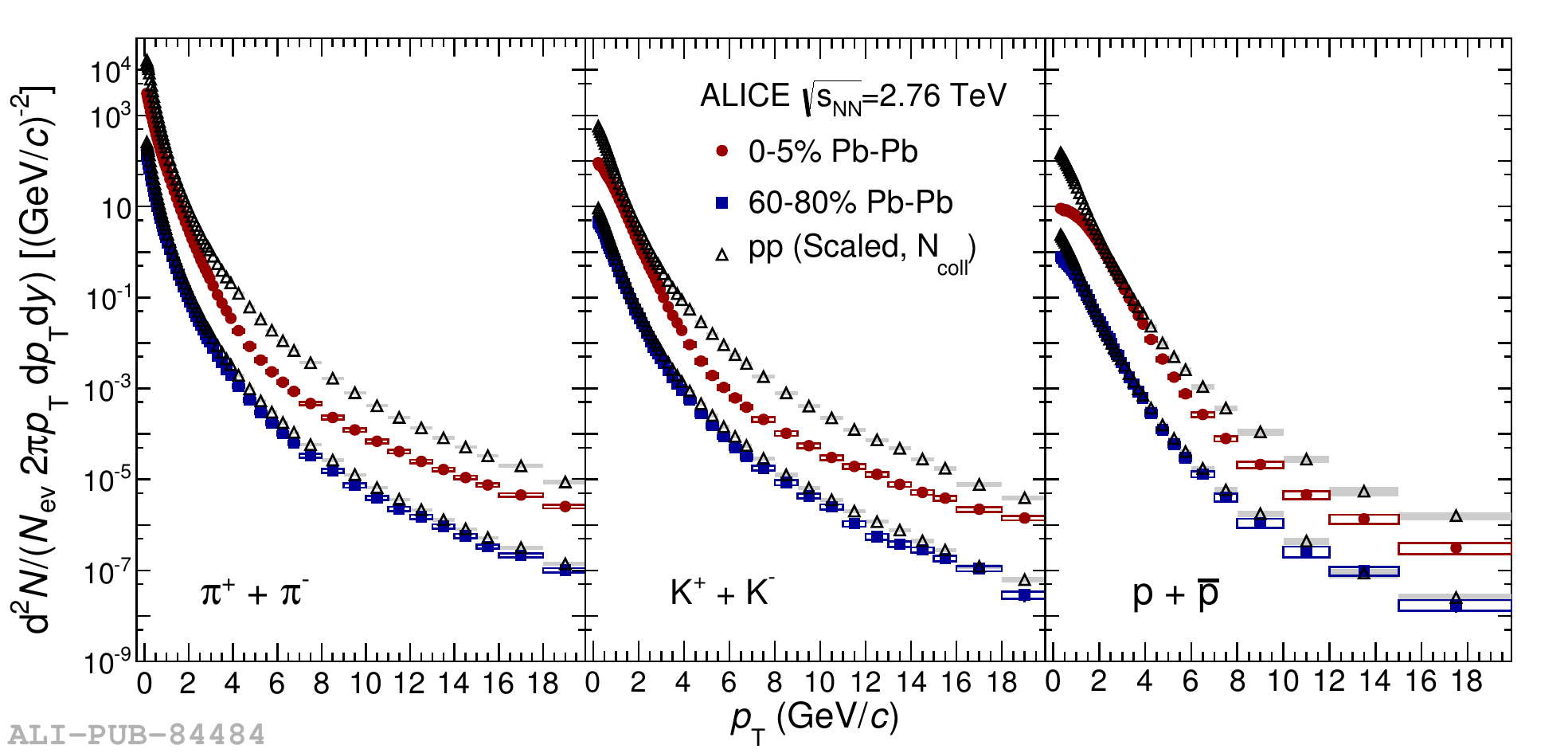}
\caption{\label{fig:pbpb:1} Solid markers show the invariant yields of identified particles in central (circles) and peripheral (squares) \pbpb collisions. Open points show the \pp reference yields scaled by the average number of binary collisions for 0-5\% (circles) and 60-80\% (squares). The statistical and systematic uncertainties are shown as vertical error bars and boxes, respectively. Figure reproduced from Ref.~\cite{Abelev:2014laa}.}
\end{center}
\end{figure*}

The system expands and cools down, when the inelastic interactions cease the yields of particles are fixed. This is the stage of the so-called chemical freeze-out which is studied using the yields of identified hadrons. As illustrated in Fig.~\ref{fig:pbpb:0}, within 20\% hadrons (except $\rm K^{*}$) are described by thermal models with a common chemical freeze-out temperature $T_{\rm ch}\approx$156 MeV. However, larger deviations are observed for protons and $\rm K^{*}$, for the latter, this is not a surprise since its mean lifetime  is comparable to that for the fireball ($\approx$10 fm/$c$)~\cite{Aamodt2011328}. As discussed here~\cite{Floris:2014pta}, the statistical model is an effective model and the small deviations from the equilibrium picture may simply indicate that the precision of the data has become sufficient to reveal its limitations.

The transverse momentum distributions of identified hadrons contain valuable information about the collective expansion of the system ($\pt\lesssim2$ GeV/$c$), the presence of new hadronization mechanisms like quark recombination ($2\lesssim \, \pt \lesssim \,8$ GeV/$c$)~\cite{Fries:2008hs} and, at larger transverse momenta, the possible modification of the fragmentation due to the medium~\cite{Sapeta:2007ad,Bellwied:2010pr}. ALICE has reported the  transverse momentum spectra, as a function of the collision centrality, of charged pions, kaons and (anti)protons from low (hundreds of MeV/$c$)~\cite{Abelev:2013vea} to high (20 GeV/$c$)~\cite{Abelev:2014laa} \pt. This is illustrated in Fig.\ref{fig:pbpb:1}, where results for \pp, the most central and the most peripheral \pbpb collisions are shown.

From the Blast-Wave  analysis~\cite{PhysRevC.48.2462} applied to the low \pt part of the spectra, the radial flow, $\langle \beta_{\rm T} \rangle$, in the most central collisions is found to be $\approx10$\% higher than at RHIC,  while the kinetic freeze-out temperature was found to be comparable to that extracted from data at RHIC, $T_{\rm kin}=$  95 MeV~\cite{Abelev:2013vea}.  The spectra are well described by hydrodynamic models, except the low \pt ($<1$ GeV/$c$) proton yield~\cite{Bozek:2012qs,Karpenko:2012yf,Werner:2012xh,Shen:2011eg}. Models which best describe the data include hadronic rescattering with non-negligible antibaryon annihilation~\cite{Werner:2012xh,Shen:2011eg}. For intermediate to high \pt ($>3$ GeV/$c$), the spectra develop the power law tail  which characterizes the hard partonic processes. 

\begin{figure*}[htbp]
\begin{center}
\includegraphics[keepaspectratio, width=1.9\columnwidth]{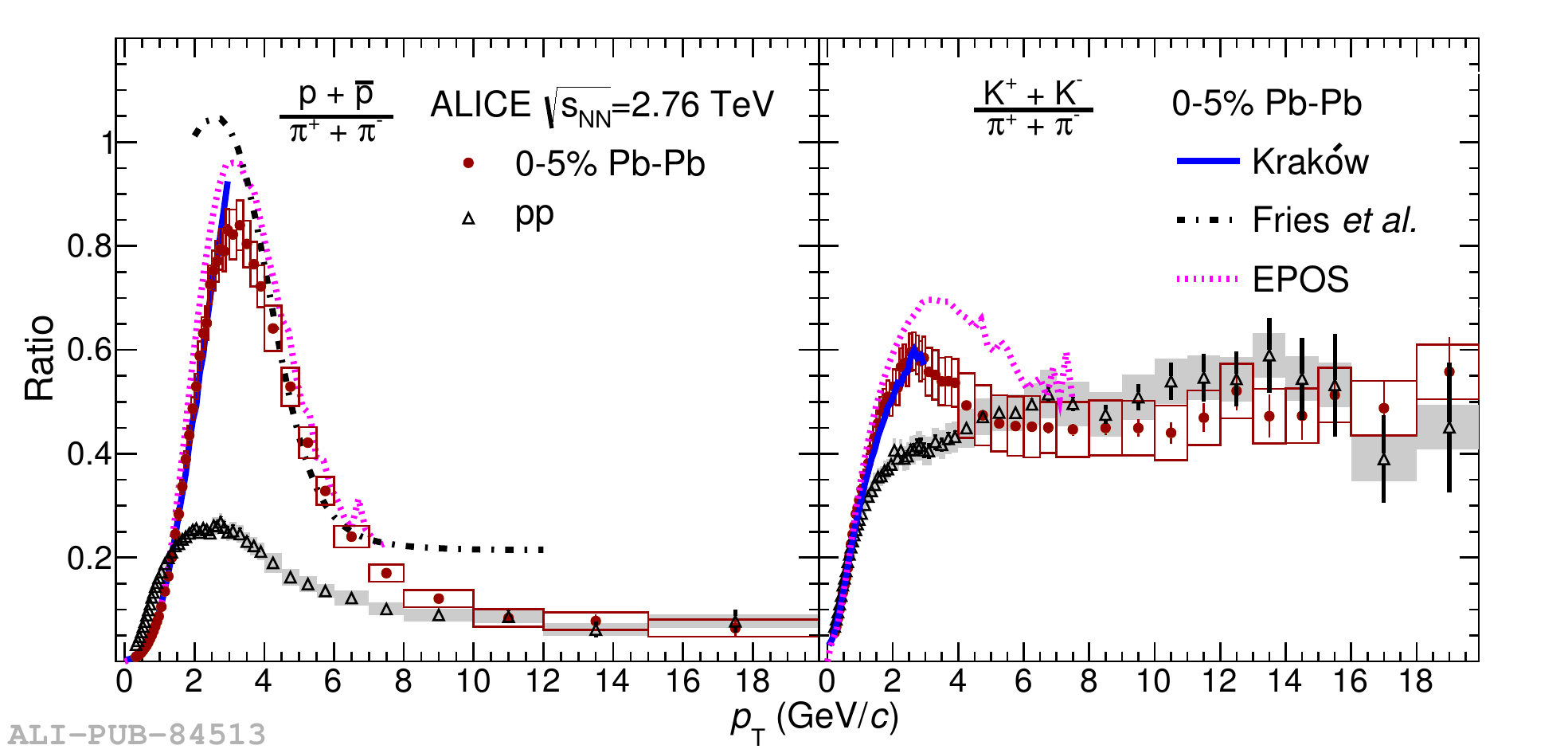}
\caption{\label{fig:pbpb:2} Particle ratios as a function of \pt measured in pp and the most central, 0-5\%, Pb-Pb collisions. Statistical and systematic uncertainties are displayed as vertical error bars and boxes, respectively. The theoretical predictions refer to \pbpb collisions. Figure reproduced from Ref.~\cite{Abelev:2014laa}.}
\end{center}
\end{figure*}

\begin{figure}[!h]
\begin{center}
  \includegraphics[height=.22\textheight]{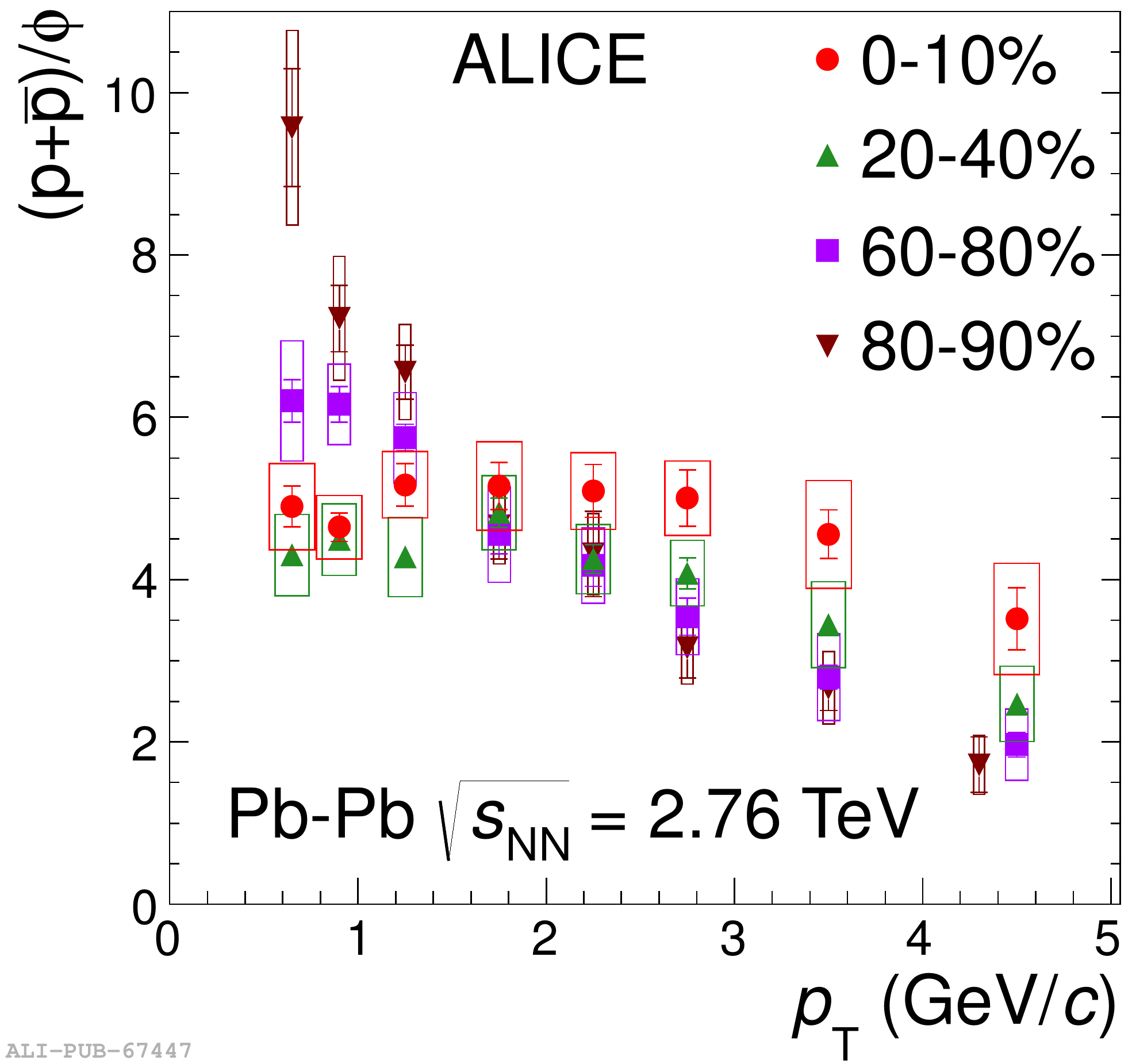}
  \includegraphics[height=.22\textheight]{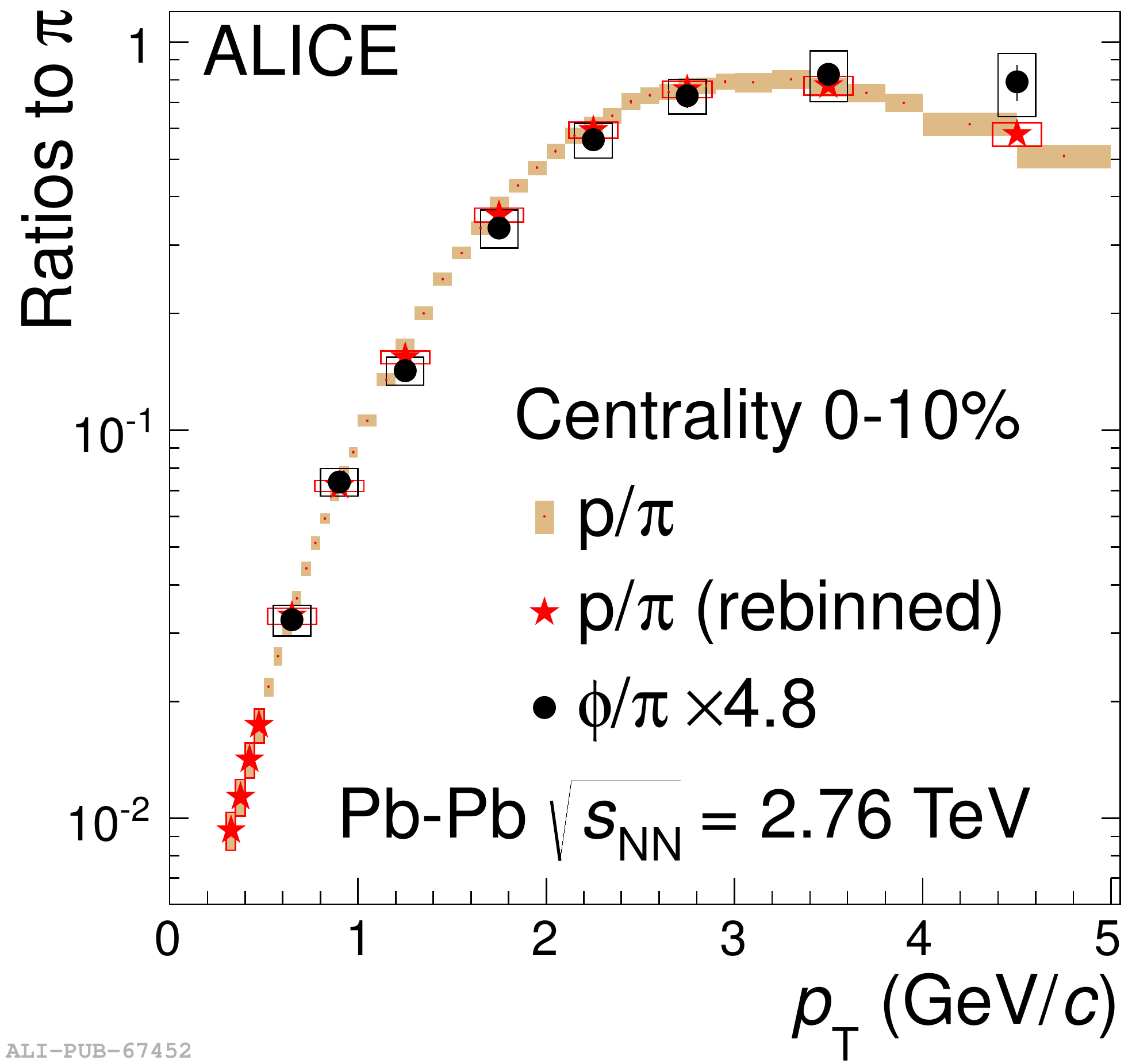}  
  \caption{Ratio p$/\phi$ as a function of \pt for \pbpb collisions at 2.76 TeV for four centrality intervals.  The statistical uncertainties are shown as bars and the total systematic uncertainties (including \pt-uncorrelated and \pt-correlated components) are shown as boxes (top). Ratios of proton and $\phi$-meson yields to charged pion yield as a function of \pt for central \pbpb collisions at 2.76 TeV.  In order to show the similarity of the shapes of the two ratios for $\pt<$3 GeV/$c$, the $\phi/\pi$ ratio has been scaled so that the $\phi$-meson and proton integrated yields are identical (bottom).}
\label{fig:pbpb:1b}
\end{center}
\end{figure}

The particle ratios as a function of \pt are shown in Fig.~\ref{fig:pbpb:2} for \pp and the most central \pbpb collisions. The proton-to-pion ratio increases from $\approx 0.38$ to $\approx 0.8$ going from peripheral (60-80\%) to central (0-5\%) \pbpb collisions at $\pt \approx 3$ GeV/$c$, then decreases to the value measured for vacuum fragmentation (\pp collisions) for $\pt > 10$ GeV/$c$. The result obtained for the most central collisions is similar to that measured at RHIC~\cite{Adare:2013esx,Abelev:2006jr}. The kaon-to-pion ratio also exhibits a bump around $\pt=$ 3 GeV/$c$, this effect is not predicted by quark recombination suggesting that the actual enhancement of the baryon-to-meson ratio is not anomalous and instead it is most likely driven by hydrodynamical flow. This picture is tested by comparing the shapes of the \pt distributions of $\phi$-meson and protons. The results shown in Fig.~\ref{fig:pbpb:1b} indicate that for central \pbpb collisions the shapes of the ratios to pions are the same. Also shown is the evolution with collision centrality of the $\phi$-meson yield normalized to that for protons as a function of \pt. For \pt below 4 GeV/$c$ the ratio become flat with the decreasing of the impact parameter. This suggests that the mass, and not the number of quark constituents, determines the spectral shape in central \pbpb collisions, this is in a good agreement with the hydro interpretation.

From the behavior of the particle ratios in both colliding systems one can establish that the suppression of the three particle species at high \pt ($>10$ GeV/$c$) is the same within the statistical and systematic uncertainties. This suggests that the chemical composition of leading particles from jets in the medium is similar to that of vacuum jets. Another result concerning jet-quenching is the first indication for the mass dependence of the parton energy-loss, this by comparing the nuclear modification factor of the prompt D mesons to that for non-prompt J$/\psi$ measured by CMS.

\begin{figure}[!h]
\begin{center}
  \includegraphics[height=.25\textheight]{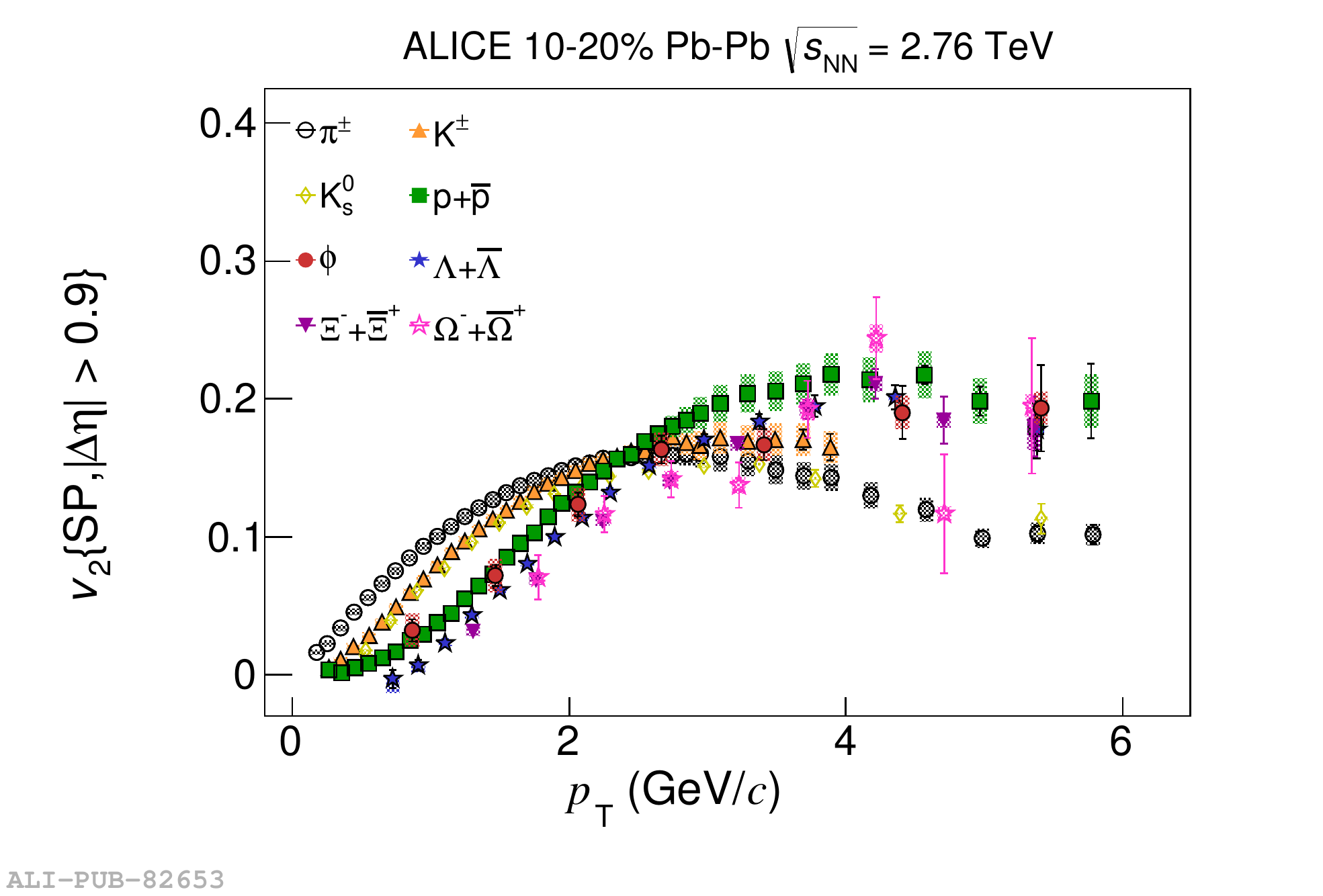}
  \includegraphics[height=.25\textheight]{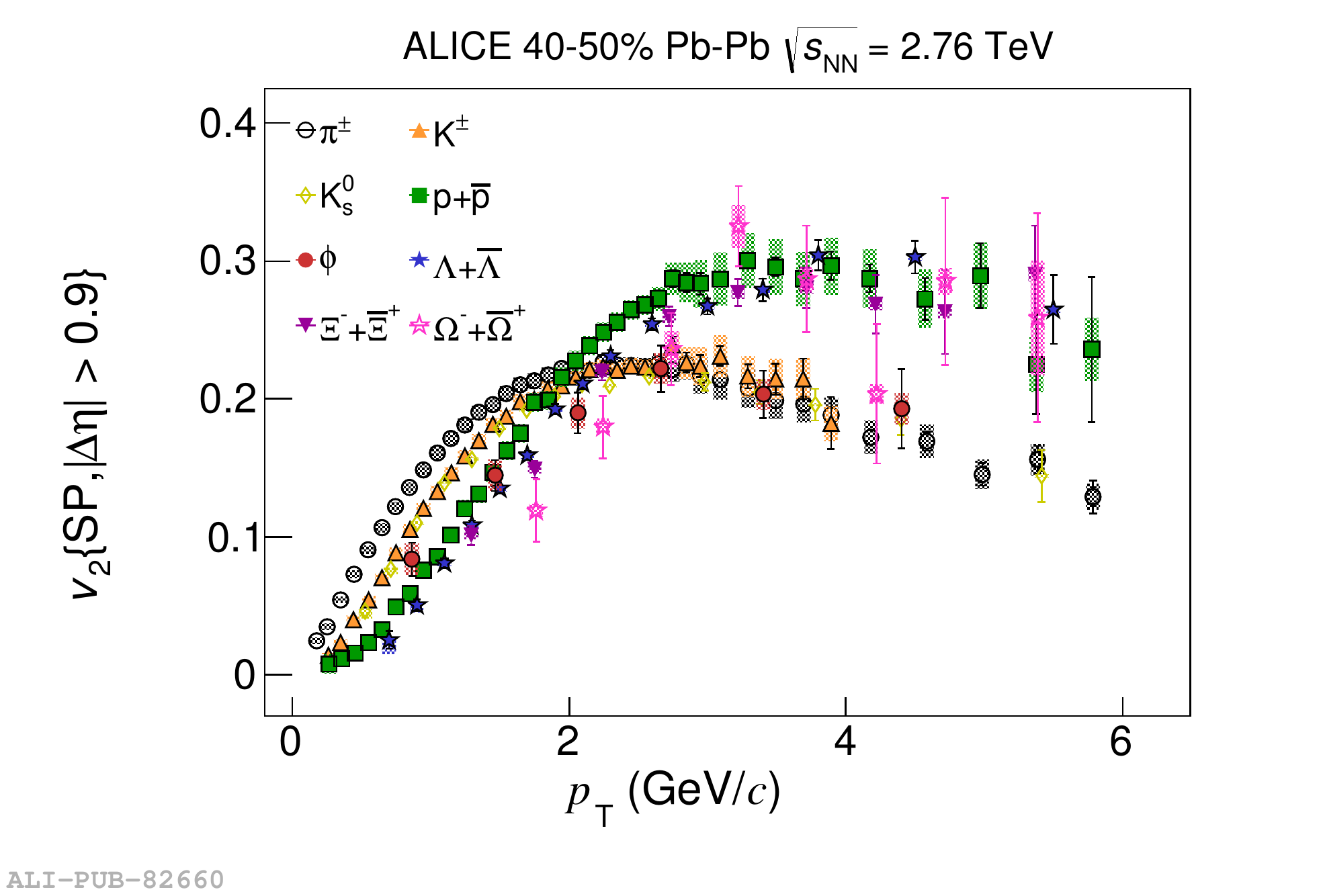}  
  \caption{Elliptic flow coefficient ($v_{2}$) of identified hadrons as a function of \pt measured for central (top) and peripheral (bottom) \pbpb collisions.}
\label{fig:pbpb:1c}
\end{center}
\end{figure}

The elliptic flow for identified hadrons has been also measured~\cite{Abelev:2014pua} over a broad \pt range. Figure~\ref{fig:pbpb:1c} shows $v_{2}$\footnote{The elliptic flow coefficients were obtained with the Scalar Product method, a two-particle correlation technique, using a pseudo-rapidity gap of $|\Delta\eta|>$0.9 between the identified hadron under study and the reference particles~\cite{Abelev:2014pua}.} as a function of \pt for central (10-20\%) and semi-peripheral (40-50\%) \pbpb collisions.  Going from central to semi-peripheral \pbpb collisions $v_{2}$ increases as expected due to the eccentricity increase. For \pt below 2 GeV/$c$ a mass ordering is observed indicating the interplay between elliptic and radial flow. For higher \pt  hadron-$v_{2}$'s seem to be grouped into baryons and mesons, the exception is the $v_{2}$ of $\phi$-mesons, which for central \pbpb collisions follows that for baryons. This observation indicates that the behavior of $v_{2}$ is driven by the hadron mass and not for the number of quark constituents. ALICE has also reported the violation of the scaling of $v_{2}$ with the number of constituent quarks, such a observation is also against the scenario with quark recombination/coalescence.  

\section{Features of the ``cold'' matter}
\label{ppb}

Surprisingly, early results from LHC showed that \ppb collisions exhibit behaviors reminiscent to those due to final state effects, namely, hints of collective effects (radial and elliptic flow, ridge structure), but no sign of jet quenching ~\cite{ABELEV:2013wsa,Abelev:2013haa}.

For \ppb collisions the transverse momentum spectra of charged pions, kaons and (anti)protons as a function of the multiplicity measured in the VZERO-A detector have been reported from hundreds of MeV/$c$~\cite{Abelev:2013haa} up to 15 GeV/$c$~\cite{Ortiz:2014iva}. At high multiplicity the low \pt parts of the spectra are better described by models which incorporate hydro~\cite{Abelev:2013haa}. Also intriguing is the fact that a similar effect has been observed in  \pp collisions, in that case the multiplicity was determined at mid-rapidity. The feature is also present in Pythia 8 tune 4C~\cite{Corke:2010yf}, this behavior is a consequence of the interactions among final partons coming from independent semi-hard scatterings which increases with increasing number of multi-parton interactions (MPI)~\cite{Ortiz:2013yxa}. One therefore cannot rule out alternative explanations, but interestingly, it illustrates that likely there is a strong coupling of this phenomenon to the underlying event also in \pp collisions.

\begin{figure*}[htbp]
\begin{center}
\includegraphics[keepaspectratio, width=1.02\columnwidth]{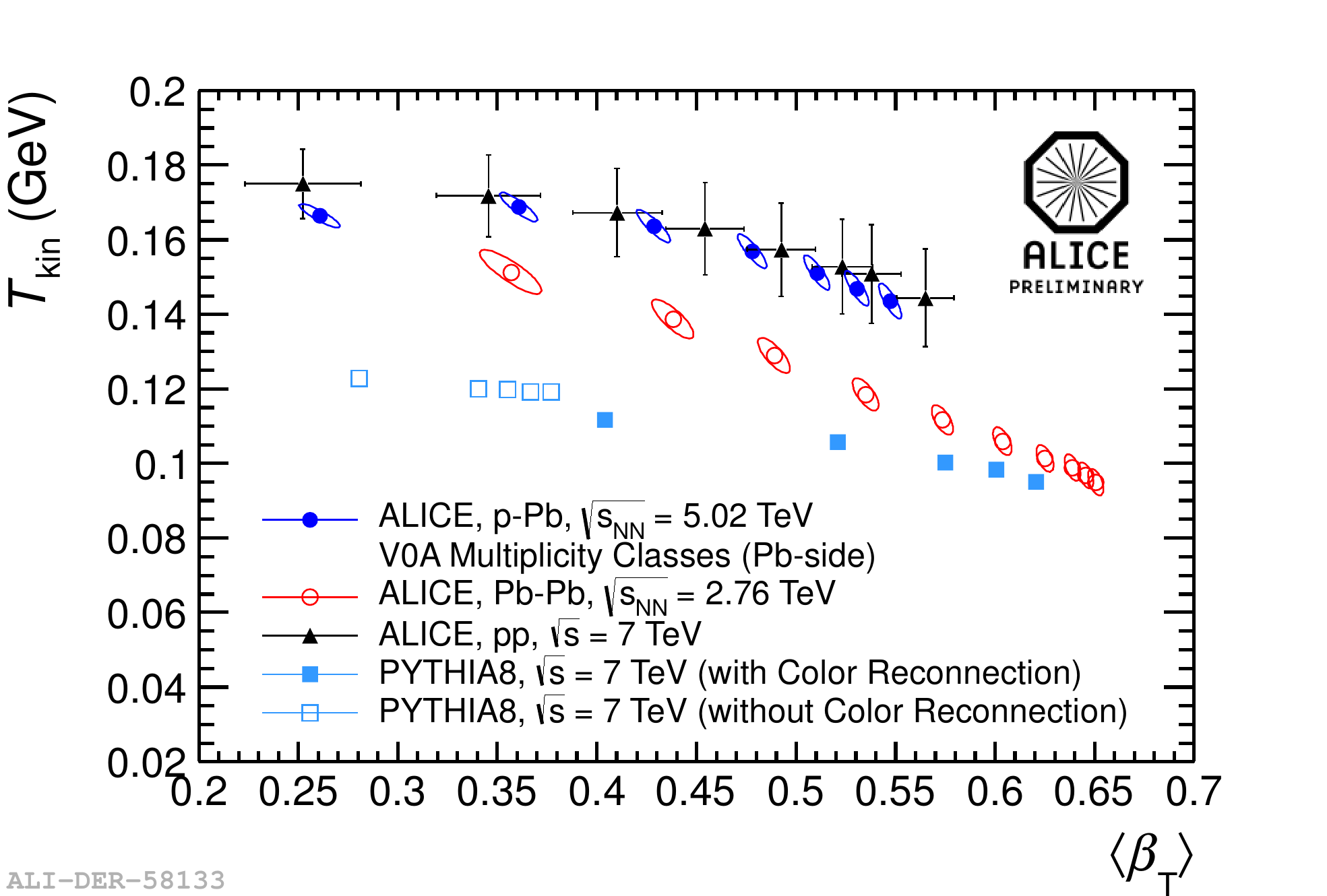}
\includegraphics[keepaspectratio, width=1.1\columnwidth]{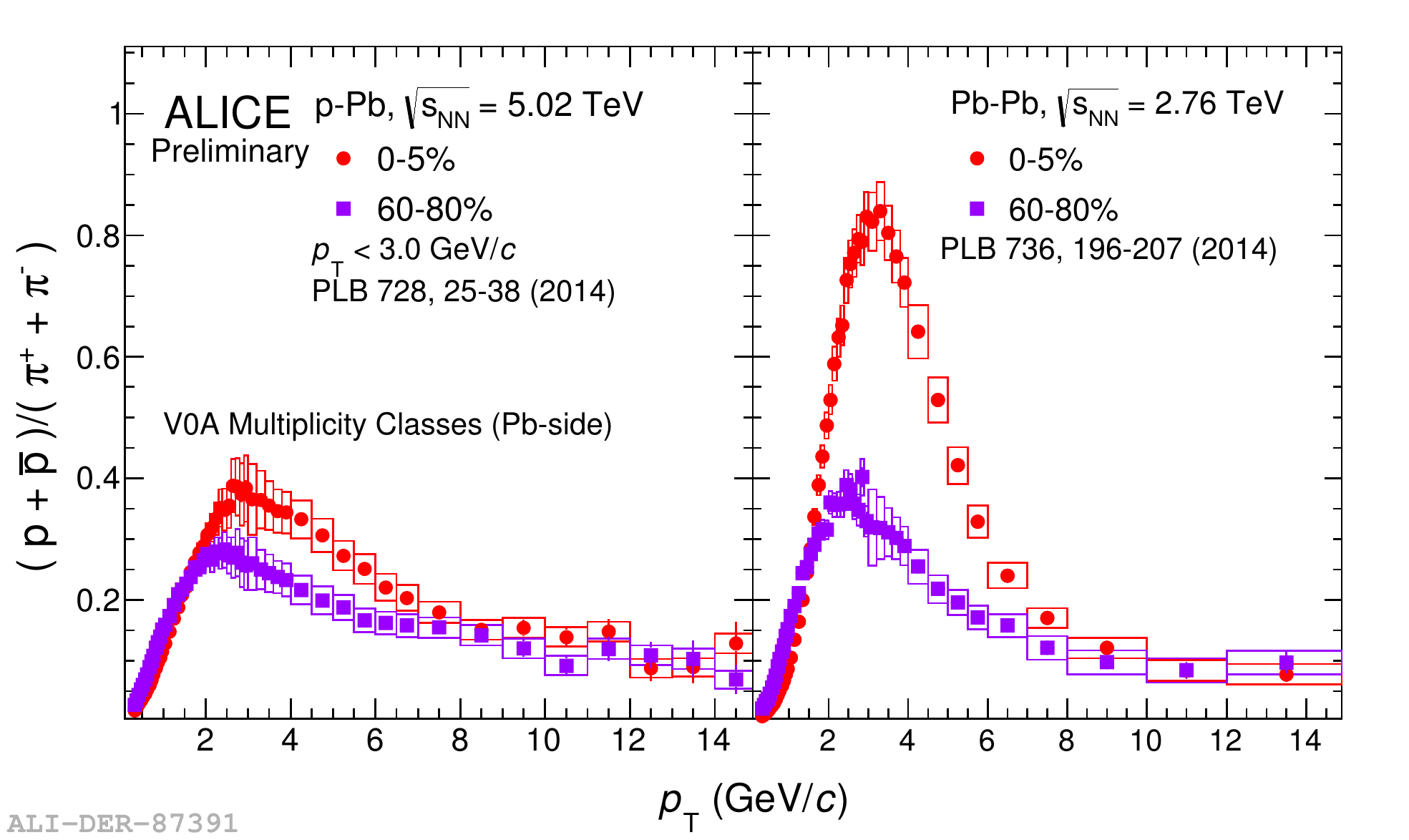}
\caption{\label{fig:pPb:1} Comparison of the results from the blast-wave analysis applied to all available systems: \pp, \ppb and \pbpb collisions. The spectral shape analysis was also applied to Pythia 8 events. Charged-particle multiplicity increases from left to right. (Right.) Proton-to-pion ratio as a function of \pt measured in \ppb and \pbpb collisions at $\sqrt{s_{\rm NN}}$ = 5.02 and 2.76 TeV, respectively. The systematic and statistical uncertainties are plotted as boxes and error bars, respectively.}
\end{center}
\end{figure*}

The evolution of the spectral shapes with multiplicity is studied using the the blast-wave analysis. Figure~\ref{fig:pPb:1} (left) shows that a qualitatively similar behavior for $T_{\rm kin}$ vs. $\langle \beta_{\rm T} \rangle$ is obtained for the three systems (\pp, \ppb and \pbpb) even in \pp events simulated with Pythia 8.  To study the effect on the \pt spectra directly, the proton-to-pion ratio was constructed, the results for \ppb and \pbpb collisions are presented in Fig.~\ref{fig:pPb:1} (right) for two extreme multiplicity intervals. For \pt below (above) 2 GeV/$c$ the ratios exhibit a depletion (enhancement) going from low  to high multiplicity. The highest (lowest) multiplicity intervals give ratios which reach maxima at $\pt \approx 3$ GeV/$c$ amounting to $\approx$0.4 and $\approx 0.8$ ($\approx 0.28$ and $\approx$ 0.38) in \ppb and \pbpb collisions, respectively. Above 3 GeV/$c$, the ratios start to decrease down to $\approx0.1$ at \pt$\approx 10$ GeV/c, which according to~\cite{Abelev:2014laa} corresponds to the value measured for vacuum fragmentation (\pp collisions).

\begin{figure*}[htbp]
\begin{center}
\includegraphics[keepaspectratio, width=1.9\columnwidth]{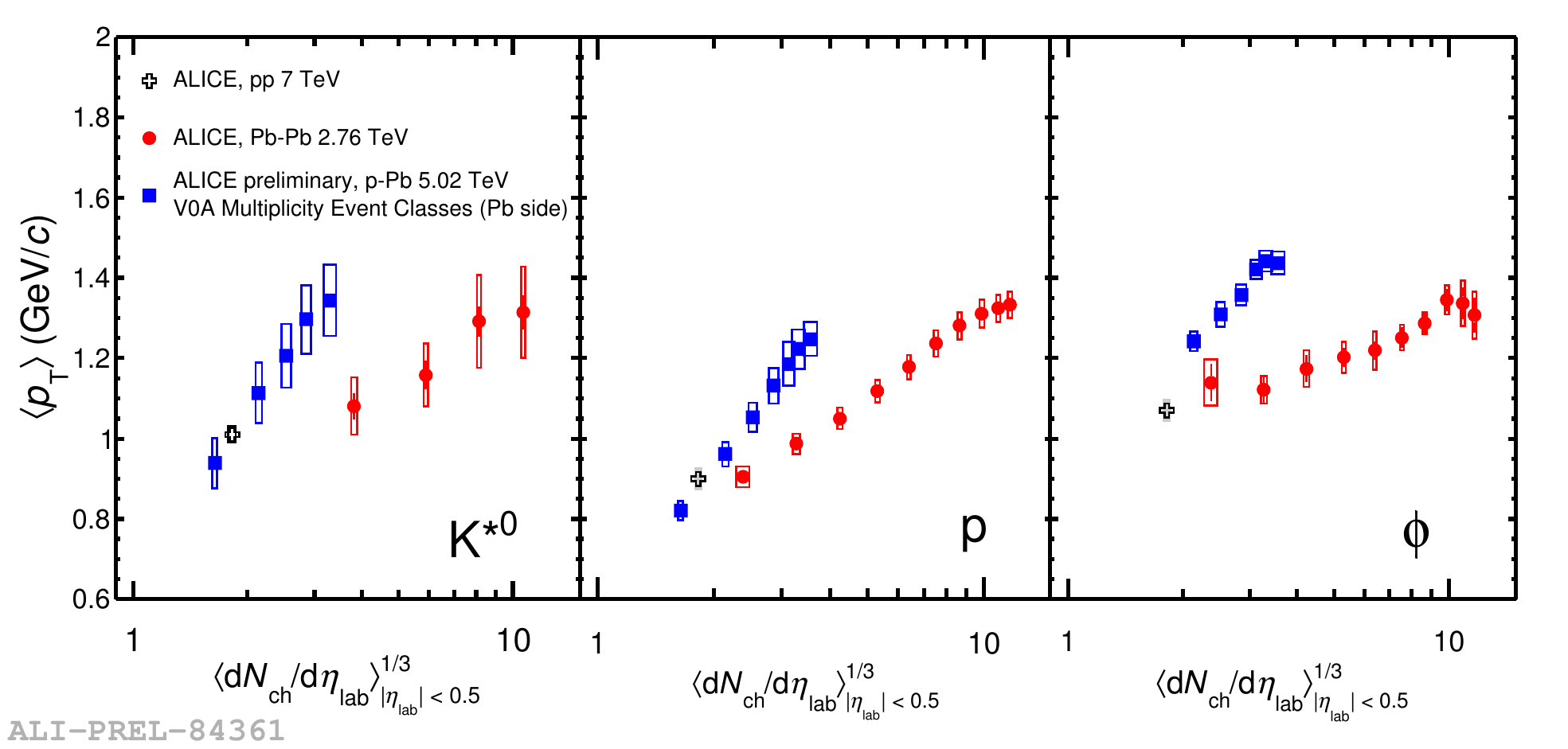}
\caption{\label{fig:pPb:2} Mean transverse momentum of resonances compared to that of proton in the three different collision systems, as a function of the system size.}
\end{center}
\end{figure*}

Figure~\ref{fig:pPb:2} shows the average \pt for ${\rm K}^{*0}$, protons and $\phi$-mesons as a function of the system size. The measurements were performed for \pp, \ppb and \pbpb collisions at $\sqrt{s_{\rm NN}}=$ 7, 5.02 and 2.76 TeV, respectively. The $\langle \pt \rangle$ for MB \pp collisions fits well with the behavior of the \ppb results. Actually, for inclusive charged particles a similar effect is seen, namely, the average \pt for \pp and \ppb data agree when the event multiplicity ($|\eta|<$0.3) is below $\sim20$, and for higher multiplicities the strongest rise of $\langle \pt \rangle$ is observed for \pp data\cite{Abelev:2013bla}. For \pbpb collisions the rise of $\langle \pt \rangle$ with the event multiplicity is weaker than for small systems. This could indicate that to reach the high multiplicity in small systems, the partonic processes need to be harder.

\begin{figure*}[htbp]
\begin{center}
\includegraphics[keepaspectratio, width=0.75\columnwidth]{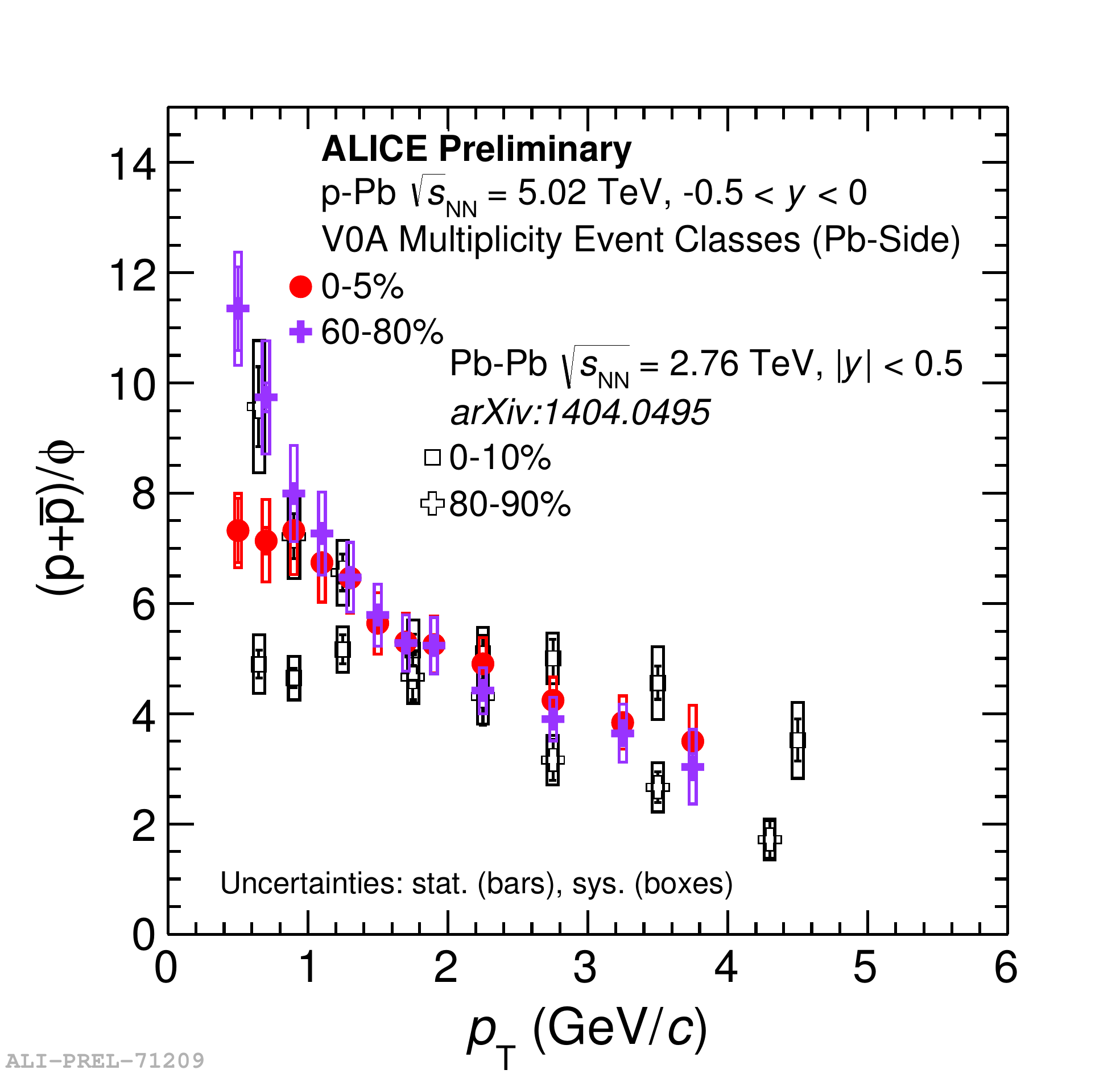}
\includegraphics[keepaspectratio, width=1.15\columnwidth]{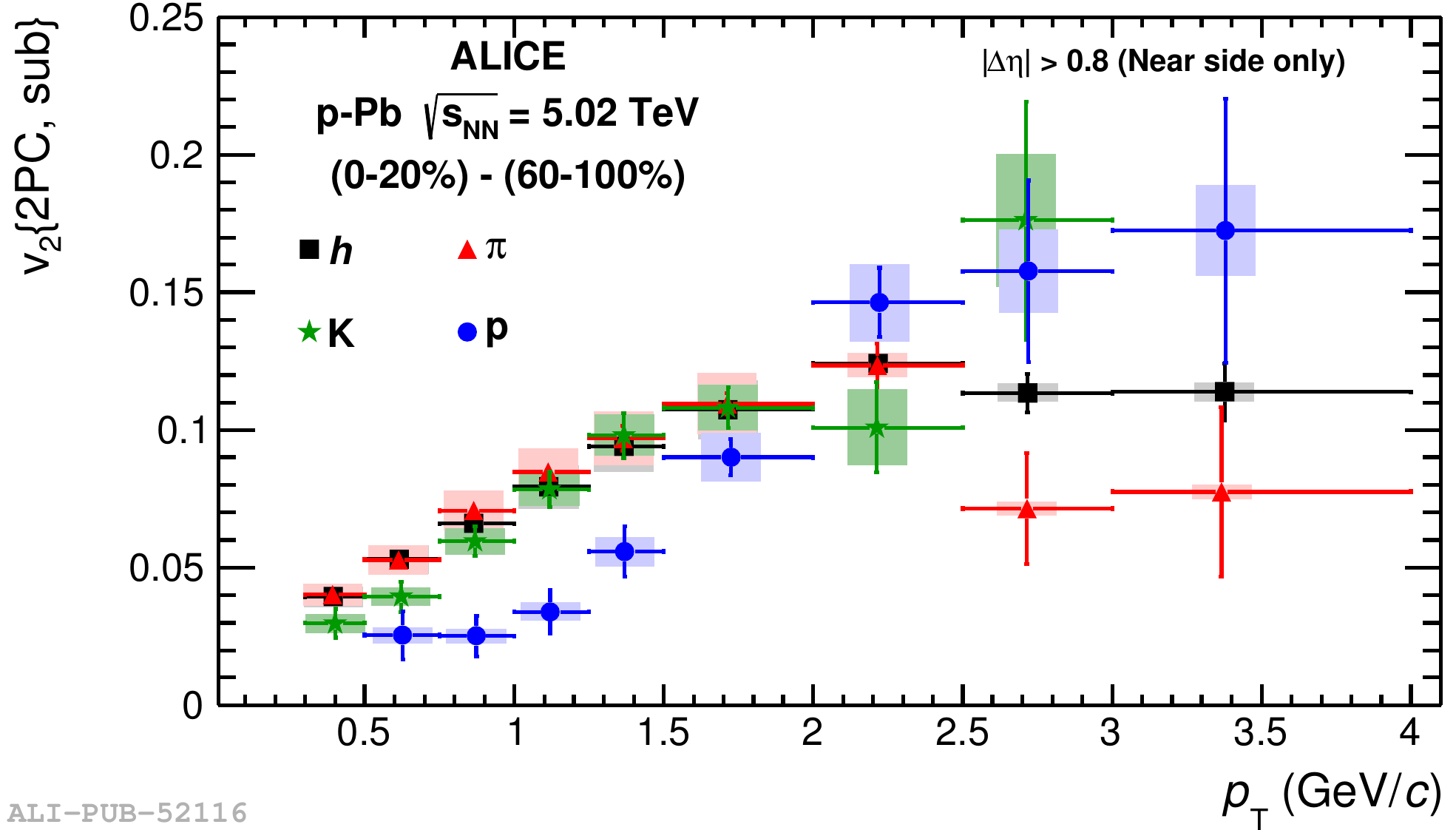}
\caption{\label{fig:pPb:3} (Left). Ratios of proton yields to $\phi$-meson yields  vs \pt in \ppb collision at 5.02 TeV and comparison with \pbpb. (Right). The Fourier coefficient $v_{2}$ for hadrons (black squares), pions (red triangles), kaons (green stars) and protons (blue circles) as a function of \pt from the correlation in the 0-20\% multiplicity class after subtraction of the correlation 	from the 60-100\% multiplicity class. Error bars show statistical uncertainties while shaded areas denote systematic uncertainties.}
\end{center}
\end{figure*}

Another interesting result is shown in Fig.~\ref{fig:pPb:3}, where the multiplicity dependence of the proton-to-$\phi$ ratio as a function of \pt is shown. The \ppb results are compared to the ratios measured in peripheral and central \pbpb collisions. The ratio for high multiplicity \ppb events exhibits a flattening for \pt below 1.5 GeV/$c$, while for higher \pt the ratio decreases in the same amount as those measured for low multiplicity \ppb and the most peripheral \pbpb collisions. The behavior of the ratio in high multiplicity \ppb data could be interpreted as a hint of the onset of collective behavior.

Another feature of the \ppb data is the non-zero elliptic flow coefficient, right panel of Fig.\ref{fig:pPb:3} shows that the \pt differential $v_{2}$ exhibits a mass ordering and a crossing for \pt around 1.5-2 GeV/$c$, again this feature is well know from heavy ion collisions. The results are reported in this paper~\cite{ABELEV:2013wsa}. To understand the potential role of MPI in \ppb collisions the minijet production as a function of the event multiplicity was studied. In \pp data this analysis was successfully applied. Through the measurement of the so-called number of uncorrelated seeds, which in Pythia are directly related with the number of MPIs, it was observed a deviation of this quantity with respect to the linear behavior, this was interpreted as an indication of the limitation of the number of MPIs to produce high multiplicity events~\cite{Abelev:2013sqa}. Contrary, in \ppb data a clear linear behavior at high multiplicity was observed, this could indicate the presence of MPIs in this colliding-systems~\cite{Abelev:2014mva}.

To close this section it is important to mention the first measurements which were published  at the LHC for high multiplicity \pp data. The CMS Collaboration reported the discovery of the ridge structure in high multiplicity events. At the same time, ALICE measured the average transverse sphericity, $S_{\rm T}$~\cite{Cuautle:2014yda}, as a function of the event multiplicity~\cite{Abelev:2012sk}, in this case it was shown that at high multiplicity $\langle S_{\rm T} \rangle$ in data exhibit an opposite behavior to those predicted by QCD-inspired models.	This can be interpreted in terms of an overestimation (by models) of the high \pt jet production. To understand the origin of these effects (flow patterns, ridge, sphericity) more detailed studies need to be done, this is one of the main topics with is being exploited in ALICE.

\section{Summary}
\label{conclusions}

In \pbpb collisions, the presented results significantly improve the precision of previous measurements in various
areas. In particular, a measurement of elliptic flow with identified particles shows a clear mass ordering for light and
strange hadrons for $\pt<$2.5 GeV/$c$. Spectra and $v_{2}$ measurements of the $\phi$ meson suggests that the mass (and not the number of constituent quarks) drives the spectral shape and the size of the elliptic flow in central collisions for $\pt<$4 GeV/$c$. While there are several observables which are approximately consistent with a description of \ppb collisions as incoherent superposition of nucleon-nucleon collisions at high \pt, some measurements hint to novel effects at low \pt which are potentially of collective origin. These findings still need to be reconciled theoretically and promise that \ppb collisions will continue to be a very exciting field in the future.

\bibliographystyle{elsarticle-num}

\bibliography{biblio}

\end{document}